\documentstyle[12pt]{ioplppt}
\begin{document}
\jl{6}

\def \D {\mbox{D}}
\def \div {\mbox{div}\,}
\def \c {\mbox{curl}\,}
\def \ep {\varepsilon}
\def \w {\widehat}
\def \ts {\textstyle}
\def \rd {\displaystyle{\cdot}}

\title{Local freedom in the gravitational field}

\author{Roy Maartens\dag\ftnote{4}{maartens@sms.port.ac.uk},
	George F R Ellis\ddag\ftnote{5}{ellis@maths.uct.ac.za} and
	Stephen T C Siklos\S\ftnote{6}{s.t.c.siklos@damtp.cam.ac.uk}}

\address{\dag\ School of Mathematical Studies, Portsmouth University,
Portsmouth PO1~2EG, Britain}

\address{\ddag\ Department of Mathematics and Applied Mathematics, 
University of Cape Town, Cape Town 7700, South Africa}

\address{\S\ Department of Applied Mathematics and Theoretical
Physics, Cambridge University, Cambridge CB3~9EW, Britain}

\begin{abstract}

In a cosmological context, the electric and magnetic parts of
the Weyl tensor,
$E_{ab}$ and $H_{ab}$, represent the locally free curvature 
 -- i.e. they are not pointwise determined by the matter
fields.                                                         
By performing a complete covariant decomposition of
$\nabla_cE_{ab}$ and $\nabla_cH_{ab}$,
we show that the parts of the derivative of the curvature
which are locally free (i.e. not pointwise determined 
by the matter via the
Bianchi identities) are exactly the symmetrised trace--free spatial
derivatives of $E_{ab}$ and $H_{ab}$ 
together with their spatial curls. These parts of the derivatives
are shown to be crucial for the existence of gravitational waves.

\end{abstract}

\pacs{0420, 9880, 9530}

\maketitle

\section{The problem}

In a cosmological context, where there is a preferred
four--velocity field $u^a$,
the electric and magnetic parts of the Weyl tensor,
$E_{ab}$ and $H_{ab}$, represent covariantly
the locally free gravitational
field, in the sense that they are not pointwise 
determined by the matter fields
(see e.g. \cite{ell71} -- \cite{veugleel96}). They 
do not arise explicitly in the Einstein equations.
Although they are not really
independent of the matter fields, 
being constrained
by integrability conditions in the form of the Bianchi
identities, it is nevertheless useful to think of them
as being {\em locally} free.
								   
The Bianchi identities involve the time derivatives and the spatial
divergences and curls\footnote{The terms {\em divergence} and
{\em curl} as applied to rank--2 tensors are defined by
equation (\ref{dc}).}
of $E_{ab}$ and $H_{ab}$ 
(see the Appendix). 
The locally free parts of the curvature
are not pointwise determined by the matter fields via 
Einstein's equations. Similarly, the locally free
parts of the derivative of the curvature are those parts not
pointwise determined via the Bianchi identities by the matter fields 
and their derivatives and by $E_{ab}$, $H_{ab}$.
The aim of this paper 
is to determine these locally free parts and  
their relation to gravitational waves in cosmology.

We find via a complete covariant decomposition
that the parts of $\nabla_cE_{ab}$ and $\nabla_cH_{ab}$
which are missing from the Bianchi identities are
exactly the projected totally symmetric and trace--free 
parts $\nabla_{\langle c}E_{ab\rangle }$ and $\nabla_{\langle c}
H_{ab\rangle }$.\footnote{Angle brackets 
$\langle \!\cdots\!\rangle $ enclosing 
indices represent their totally symmetric trace-free and spatially
projected part. Round and square brackets denote, respectively,
symmetrisation and anti--symmetrisation.} We call these parts
the {\em distortions}
of $E_{ab}$ and $H_{ab}$.
The distortions
represent the part of the
derivative of the curvature which 
is uncoupled from the Bianchi identities and
does not locally interact with the matter fields. 
The divergences are determined pointwise by the matter terms
and the locally free field.
The curls (together with the matter terms and the locally free field) 
determine the time derivatives of $E_{ab}$ and $H_{ab}$ 
pointwise. Therefore  
{\em the distortions and the curls
 provide a covariant characterisation of
the locally free part of the space--time gradient
of the gravitational field.} 

We show that in the case of gravitational waves (covariantly
characterised), the Laplacian is determined by the 
distortion. Hence
{\em not only the curls, but also the distortions of 
$E_{ab}$ and $H_{ab}$ must be non--zero if there are
gravitational waves.} This result represents a 
refinement of the covariant characterisation and understanding of
gravitational waves in cosmology
(see e.g. \cite{h} -- \cite{dbe}).

In Section 2 we give the complete
covariant decompositions of the derivatives
of the Weyl tensor and fluid kinematic quantities. The 
derivations of these
decompositions, involving a combination of 
covariant tensor methods and Young diagram methods, 
are presented in Section 3. In Section 4 we show how 
the distortions and curls determine
the covariant gravitational wave equations.
Further implications of our results
are discussed 
in Section 5. The Appendix gives the
Ricci and Bianchi identities in $1+3$ covariant form,
together with the required identities for second order derivatives.

\section{The basic results}

The locally free field $E_{ab}$, $H_{ab}$ and 
its characterisation are covariant if a covariantly defined
four--velocity exists. For a single perfect fluid,
there is a unique covariant four--velocity,
whereas for imperfect fluids and
multi--fluids there is in general more than one covariantly
defined four--velocity. 
For simplicity, we restrict ourselves
to the perfect fluid case. The results are readily extended
to the imperfect fluid or multi--fluid case, given any 
covariant choice of four--velocity.

For a gravitational field with perfect fluid source, 
the basic covariant variables are: the
fluid scalars $\Theta$ (expansion), $\rho$ (energy density),
$p$ (pressure); the fluid spatial vectors $\dot{u}_a$ 
(4--acceleration),
$\omega_a$ (vorticity); the spatial trace--free symmetric tensors
$\sigma_{ab}$ (fluid shear), $E_{ab}$, $H_{ab}$; and the projection
tensor $h_{ab}=g_{ab}+u_au_b$, which 
projects orthogonal to the fluid 
4--velocity vector $u^a$. 
The covariant spatial derivative is defined by
\[ 
\D_c S^{a\cdots}{}{}_{b\cdots}=h_c{}^rh^a{}_p\cdots h_b{}^q\cdots
\nabla_rS^{p\cdots}{}{}_{q\cdots}\,.
\]
It follows that $\D_ch_{ab}=0$.
The covariant projected permutation tensor is
$\ep_{abc}=\eta_{abcd}u^d$, satisfying
$\D_a\ep_{bcd}=0$ and
$\ep_{abc}\ep^{def}=3!h_{[a}{}^dh_b{}^eh_{c]}{}^f$.
The covariant spatial divergence and curl of rank--2 tensors 
are defined by \cite{maart96}
\begin{equation}
(\div S)_a=\D^b S_{ab}\,,\quad (\c S)_{ab}=
\ep_{cd(a}\D^c S_{b)}{}^d \,,
\label{dc}\end{equation}
generalising the expressions for vectors:
\[
\div V =\D^bV_b\,,\quad(\c V)_a=\ep_{abc}\D^b V^c\,.
\]
For brevity, we shall henceforth omit the brackets around
`curl' unless this leads to ambiguity.

The aim now is, through the covariant $1+3$ 
splittings of vectors and rank--2 and
rank--3 tensors, to find the complete covariant decomposition 
of the derivatives 
of scalars, vectors, and rank--2 symmetric tensors.
Any space--time vector $S_{a}$ may be covariantly
split as:
\[
S_a = - S_b u^b u_a + h_a{}^c S_{c}\,.
\]
As a special case, the derivatives of
the fluid scalars have the decomposition
\[
\nabla_a\Theta=-\dot{\Theta}u_a+\D_a\Theta\,,
\]
and similarly for $\rho$ and $p$.

Any spatial tensor $S_{ab}=h_a{}^ch_b{}^dS_{cd}$ may be covariantly
decomposed as:
\begin{equation}
S_{ab}=S_{\langle ab\rangle }+{\ts{1\over3}}h_{cd}S^{cd}
h_{ab}+\ep_{abc}S^c\,,
\label{0}\end{equation}
where $ S_{\langle ab\rangle }\equiv h_{(a}{}^ch_{b)}{}^d S_{cd}-
{\ts{1\over3}}h_{cd}S^{cd}h_{ab} $
is the projected symmetric trace--free part, and the spatial vector
$S^a$ is equivalent to the skew part:
\begin{equation}
S_{[ab]}=\ep_{abc}S^c\quad\Leftrightarrow\quad
S_a={\ts{1\over2}}\ep_{abc}S^{bc}\,.
\label{0''}\end{equation}
If $S_{ab}=S_{(ab)}$, then $(\c S)_{ab}=(\c S)_{\langle ab\rangle }$. 
If $S_{ab}=
S_{[ab]}$, then we find the identities
\begin{equation}
\c S_{ab}=\D_{\langle a}S_{b\rangle }-{\ts{2\over3}}
\D^cS_ch_{ab}\,,\quad
\D^bS_{ab}=\c S_a\,.
\label{0'}\end{equation}

The decomposition of the derivative of
the fluid 4--velocity \cite{ell71} is in our notation
\begin{eqnarray}
&& \nabla_bu_a=-\dot{u}_au_b+\D_bu_a\,, \label{0a} \\
&&\D_bu_a=\sigma_{ab}+{\ts{1\over3}}\Theta h_{ab}
+\omega_c\ep_{ab}{}{}^c\,, \label{3}\\
&& \sigma_{ab}=\D_{\langle a}u_{b\rangle }\,,~\Theta=\D^a u_a\,,
~\omega_a=-
{\ts{1\over2}}\c u_a\,, \label{0c}
\end{eqnarray}
where (\ref{3}) is a particular case of (\ref{0}).  For the vorticity,
using (\ref{0}) -- (\ref{0'}) we find
\begin{eqnarray}
&& \nabla_b\omega_a=-\omega_c\dot{u}^cu_au_b-
\dot{\omega}_{\langle a\rangle }u_b
-u_a\left\{\sigma_{bc}\omega^c+{\ts{1\over3}}\Theta\omega_b\right\}
+\D_b\omega_a\,, \label{0d} \\
&&\D_b\omega_a=\D_{\langle a}\omega_{b\rangle }+
{\ts{1\over3}}\D^c\omega_c h_{ab}
-{\ts{1\over2}}\c\omega_c\ep_{ab}{}{}^c\,,
 \label{0e}
\end{eqnarray}
where $\dot{S}_{\langle a\rangle }\equiv h_a{}^b\dot{S}_b$,
and similarly for the 4--acceleration $\dot{u}_a$.

The electric and magnetic parts of the Weyl tensor,
$ E_{ab}=C_{acbd}u^cu^d$ and $H_{ab}={\ts{1\over2}}
\ep_{acd}C^{cd}{}{}_{be}u^e$,
satisfy $E_{ab}=E_{\langle ab\rangle }$ 
and $H_{ab}=H_{\langle ab\rangle }$.
Their derivatives have the following covariant 
decomposition: 
\begin{eqnarray}
\nabla_cE_{ab} &=& -\dot{E}_{\langle ab\rangle }u_c-2u_{(a}E_{b)d}
\dot{u}^du_c \nonumber\\
&&{}+2u_{(a}\left\{E_{b)}{}^d\left(\sigma_{dc}+\omega_{dc}\right)
+{\ts{1\over3}}\Theta E_{b)c}\right\} +\D_cE_{ab} \,,\label{1} \\
\D_cE_{ab} &=& \w{E}_{cab}+
{\ts{3\over5}} \D^dE_{d\langle a}h_{b\rangle c} 
-{\ts{2\over3}} \c E_{d(a}\ep_{b)c}{}{}^d \,,
\label{2} 
\end{eqnarray}
where $\dot{S}_{\langle ab\rangle }\equiv 
h_{(a}{}^ch_{b)}{}^d\dot{S}_{cd}-
{1\over3}h_{cd}\dot{S}^{cd}h_{ab}$, and
where $\w{E}_{cab}$ is completely symmetric and 
trace--free:
\begin{equation}
\w{E}_{cab}=\w{E}_{\langle cab\rangle } =\D_{\langle c}
E_{ab\rangle }\,. \label{2a}
\end{equation}
Note that the projected time derivative in (\ref{1}) satisfies
$h_a{}^dh_b{}^e\dot{E}_{de}=h_{\langle a}{}^dh_{b\rangle }{}^e
\dot{E}_{de}$.
This decomposition is derived in the next section.            \\

A similar decomposition applies to $\nabla_cH_{ab}$. 
The generalisation (\ref{2}) of (\ref{3}) 
suggests that we call
$\w{E}_{cab}$ and $\w{H}_{cab}$ the 
distortions of $E_{ab}$ and $H_{ab}$, by analogy with
the shear $\sigma_{ab}$. These represent the 
divergence--free and curl--free spatial variation of the Weyl
tensor.\footnote{Note that the term
distortion in the case of a rank--2 tensor is not intended to
imply the geometrical significance that it has for
a vector \cite{ell71}.}
The decomposition (\ref{1}) -- (\ref{2a}) applies to
the derivative of any spatial, symmetric and trace--free rank--2
tensor;
thus we have for example
\begin{equation}
\D_c\sigma_{ab} = \w{\sigma}_{cab}+
{\ts{3\over5}} \D^d\sigma_{d\langle a}h_{b\rangle c} 
-{\ts{2\over3}} \c \sigma_{d(a}\ep_{b)c}{}{}^d \,,
\label{2e} 
\end{equation}
in the case of the shear $\sigma_{ab}$ .
 
Using the notation of \cite{maart96}, the 
covariant splitting of the
Ricci and Bianchi identities given in \cite{ell71}
is considerably simplified. For convenience, these equations
are given in the appendix, in the case of a perfect fluid source
for the gravitational field. All the covariant time and spatial
derivatives of the scalar and vector variables --
$\Theta$, $\rho$, $p$, $u_a$, $\omega_a$ and $\dot{u}_a$ --
appear in the covariant Ricci and Bianchi
equations. This includes the distortions 
$\D_{\langle b}u_{a\rangle }\equiv
\sigma_{ab}$, $\D_{\langle b}\omega_{a\rangle }$ and 
$\D_{\langle b}\dot{u}_{a\rangle }$. 
But for the tensor variables $E_{ab}$, 
$H_{ab}$ and $\sigma_{ab}$, the distortions do not appear in
any of these equations. 
In the case of the shear, the divergence 
and curl terms occur in the 
Ricci constraint equations (\ref{a4}) and (\ref{a6}) respectively,
but the distortion $\w{\sigma}_{cab}$
does not occur in any of the Ricci or Bianchi identities. 
However, the evolution of $\w{\sigma}_{cab}$ is governed by
the distortion of $E_{ab}$, as may be seen by taking the 
symmetric and trace--free part of the spatial derivative of the
shear propagation equation (\ref{a3}).

The distortion arises in the covariant Laplacian,
which is central to the wave equation
(see Section 4). Consider the distortion of say $H_{ab}$:
\[
\w{H}_{cab}=\D_{(c}H_{ab)}-{\ts{2\over5}}h_{(ab}\D^dH_{c)d}\,.
\]
Taking the divergence, we get
\[
\D^c\w{H}_{cab}={\ts{1\over3}}\D^2H_{ab}-{\ts{4\over15}}\D_{\langle a}
\D^cH_{b\rangle c}+{\ts{2\over3}}\D^c\D_{\langle a}H_{b\rangle c}\,,
\]
where $\D^2=\D^a\D_a$ is the covariant Laplacian.
The last term on the right may be converted into a divergence term
plus curvature correction terms, using the commutation identity
(\ref{a13}):
\begin{equation}
\D^2H_{ab} = 3\D^c\w{H}_{cab}
-2\left(\rho-{\ts{1\over3}}\Theta^2\right)H_{ab}
-{\ts{6\over5}}\D_{\langle a}\D^c H_{b\rangle c}
+{\cal N}[H]_{ab}\,,
\label{lap}\end{equation}
where
\begin{eqnarray}
{\cal N}[H]_{ab} &=& 4\omega^c\omega_cH_{ab}+2\Theta
\sigma^c{}_{\langle a}H_{b\rangle c}
-6\omega^c\omega_{\langle a}H_{b\rangle c} \nonumber\\
&& +2\sigma_{c\langle a}\sigma_{b\rangle d}H^{cd}-
2\sigma^{cd}H_{cd}\sigma_{ab}
+2\Theta\omega^c\ep_{cd\langle a}H_{b\rangle }{}^d \nonumber\\
&&-2\sigma^{cd}\sigma_{c\langle a}H_{b\rangle d}- 
6E^c{}_{\langle a}H_{b\rangle c}
+2\omega^c\ep_{cde}\sigma^d{}_{\langle a}H_{b\rangle }{}^e \nonumber\\
&&+2\omega^c
\ep_{cd\langle a}\sigma_{b\rangle e}H^{de}
 +2\omega_c\ep^{cd}{}{}_a\dot{H}_{\langle bd\rangle }
 +2\omega_c\ep^{cd}{}{}_b\dot{H}_{\langle ad\rangle }\,.
\label{lapn}\end{eqnarray}
The new identity (\ref{lap}), (\ref{lapn}) 
shows explicitly how the covariant
Laplacian is determined by the divergence of the distortion, 
the gradient
of the divergence, the time derivative, and algebraic terms. 
The homogeneous and isotropic Friedman-Robertson-Walker (FRW)
universe is covariantly characterised by the vanishing of $\omega_a$,
$\dot{u}_a$ and $\sigma_{ab}$, which then forces $E_{ab}$, $H_{ab}$
and the spatial gradients of $\rho$, $p$ and $\Theta$ to vanish. In
an almost FRW universe (which is considered in Section 4), 
these covariant quantities are 
small \cite{bruell93}, and the term ${\cal N}[H]_{ab}$ is non--linear
and may be neglected.

\section{The decompositions}

The derivations of the 
covariant decompositions (\ref{0a}) -- (\ref{0e})
for the vectors $u_a$ and $\omega_a$ (and the corresponding 
expressions for $\dot{u}_a$) involve a straightforward application
of the $1+3$ splitting of rank--2 tensors together with (\ref{0}) --
(\ref{0'}). For the tensor derivatives, we need the $1+3$
splitting of rank--3 tensors, the generalisation to rank--3 tensors of
the decomposition (\ref{0}), and the properties of the
covariant tensor curl and divergence.

\noindent{\bf A: The space-time splitting.} 
Any rank--3 space--time tensor has the covariant $1+3$ decomposition
\begin{eqnarray*}
S_{abc} &=& \alpha u_au_bu_c+V_au_bu_c+W_bu_cu_a+Z_cu_au_b \\
&& +A_{ab}u_c+B_{bc}u_a+C_{ca}u_b+F_{abc} \,,
\end{eqnarray*}
where $V_a,\cdots,F_{abc}$ are spatial tensors. Since
$E_{ab}=E_{(ab)}$ and $E_{ab}u^b=0$,
we have $u^au^bE_{ab;c}=0$ and so
\[
E_{ab;c}=2V_{(a}u_{b)}u_c+A_{ab}u_c+2u_{(a}B_{b)c}+F_{abc} \,,
\]
where $A_{ab}=A_{(ab)}$ and 
$F_{abc} =\D_cE_{ab}$. Contracting
with $u^c$ and $u^b$ and using (\ref{3}) gives
\begin{eqnarray*}
A_{ab}+V_au_b+V_bu_a &=& -\dot{E}_{ab}\,,\\
B_{ac}+ V_au_c &=&\left({\ts{1\over3}}\Theta h^b{}_c
+\sigma^b{}_c+\omega^b{}_c-\dot{u}^bu_c\right)E_{ab}\,,
\end{eqnarray*}
and (\ref{1}) follows.

\noindent{\bf B: The spatial splitting}. 
Any rank--3 tensor in $n$ dimensions may be decomposed as
\begin{equation}
M_{abc}=M_{(abc)}+S_{abc}+T_{abc}+M_{[abc]}\,,
\label{s1}\end{equation}
where
\begin{eqnarray}
S_{abc} &=& {\ts{2\over3}}M_{[ab]c}+{\ts{2\over3}}M_{[a|c|b]}=
-S_{bac}\,,\nonumber\\
T_{abc} &=& {\ts{2\over3}}M_{a[bc]}+{\ts{2\over3}}M_{[b|a|c]}=
-T_{acb}\,.\label{s2}
\end{eqnarray}
In terms of Young diagrams \cite{sik96}, this is represented by
the irreducible decomposition
$$
{\em
\begin{tabular}{|c|}\hline
a \\ \hline
\end{tabular}
\otimes
\begin{tabular}{|c|}\hline
b \\ \hline
\end{tabular}
\otimes
\begin{tabular}{|c|}\hline
c \\ \hline
\end{tabular}
=
\begin{tabular}{|c|c|c|}\hline
a&b&c \\ \hline
\end{tabular}
\oplus \begin{tabular}{|c|c}     \hline
b&\multicolumn{1}{c|}{c} \\ \hline
a&{} \\ \cline{1-1}
\end{tabular}
\oplus \begin{tabular}{|c|c}\hline
b& \multicolumn{1}{c|}{a}\\ \hline
c&{} \\ \cline{1-1}
\end{tabular}
\oplus \begin{tabular}{|c|}\hline
a \\ \hline
b \\ \hline
c \\ \hline
\end{tabular} \,,
}
$$
so the 4 terms on the right of (\ref{s1}) are linearly independent.
For $n=3$, they have $10+8+8+1=27$ independent components. 
(In general, they have
\begin{eqnarray*}
&&{\ts{1\over3!}}n(n+1)(n+2)+{\ts{1\over3}}n(n+1)(n-1)+{\ts{1\over3}}
n(n+1)(n-1)\\
&&+{\ts{1\over3!}}n(n-1)(n-2)=n^3
\end{eqnarray*}
free components.) Since the terms in (\ref{s1}) are linearly 
independent, we can split off their trace--free parts
independently. Because of the symmetry, each has (at most) one trace
term. Using the appropriate metric $h_{ab}$:
\begin{eqnarray}
M_{(abc)}&=&M_{\langle abc\rangle }+{3\over n+2}M_{(a}h_{bc)}\,,
\label{s3}\\
S_{abc}&=& \w{S}_{abc}+{2\over n-1}S_{[a}h_{b]c}\,,\label{s4}\\
T_{abc}&=& \w{T}_{abc}+{2\over n-1}T_{[b}h_{c]a}\,,\label{s5}
\end{eqnarray}
where $M_a=M_{(abc)}h^{bc}$, $S_a=S_{abc}h^{bc}$, $T_a=T_{abc}h^{ac}$,
and $\w{S}_{abc}$, $\w{T}_{abc}$ are trace--free.
If $M_{abc}=M_{(ab)c}$, then symmetrising (\ref{s1}) on $ab$ gives
\begin{equation}
M_{abc}=M_{(abc)}+T_{(ab)c}\,,
\label{s6}\end{equation}
and (\ref{s2}) becomes
\begin{equation}
T_{abc}={\ts{4\over3}}M_{a[bc]}\,.
\label{s7}\end{equation}
Notice that, when $M_{abc}=M_{bac}$, anti--symmetrising (\ref{s6})
on $bc$ gives, using (\ref{s7}),
${\ts{3\over2}}T_{abc}=T_{(ab)c}-T_{(ac)b}$,
so $T_{(ab)c}$ is equivalent to $T_{abc}$. As before, we can
split off the trace--free parts of (\ref{s6}).
If $M_{abc}=M_{[ab]c}$, then $T_{[ab]c}=0$ and
(\ref{s1}) becomes
\begin{equation}
M_{abc}=S_{abc}+M_{[abc]}\,.
\label{s8}\end{equation}

Specialising now to the case of {\it spatial rank--3 tensors},
so that $n=3$, simplifications result from (\ref{0''}):
\begin{equation}
M_{[abc]}= M\ep_{abc}\,,~
\w{S}_{abc}=\w{S}_{dc}\ep_{ab}{}{}^d\,,~
\w{T}_{abc}=\w{T}_{da}\ep_{bc}{}{}^d\,,
\label{s9}
\end{equation}
where $\w{S}_{cd}={\ts{1\over2}}\w{S}_{abc}\ep^{ab}{}{}_d
=\w{S}_{\langle cd\rangle }$ and similarly for $\w{T}_{cd}$.
If $M_{abc}=M_{(ab)c}$, then by (\ref{s3}), (\ref{s5}) -- (\ref{s7})
and (\ref{s9}), we have
\begin{equation}
M_{abc}\equiv M_{(ab)c}=M_{\langle abc\rangle }+{\ts{3\over5}}
M_{(a}h_{bc)}+
{\ts{1\over2}}\w{T}_{d(a}
\ep_{b)c}{}{}^d+{\ts{1\over2}}T_{[b}h_{c]a}+T_{[a}h_{c]b}\,,
\label{s12}\end{equation}
where $M_a={\ts{2\over3}}M_{ad}{}{}^d-{\ts{1\over3}}M_d{}^d{}{}_a$
and $T_a={\ts{2\over3}}M_{ad}{}{}^d-{\ts{2\over3}}M_d{}^d{}{}_a$.
If in addition  $M_a{}^a{}{}_c=0$, then $M_a=T_a$ and (\ref{s12})
becomes
\begin{equation}
M_{abc}\equiv M_{\langle ab\rangle c}=M_{\langle abc\rangle }+
{\ts{9\over10}}M_{\langle a}h_{b\rangle c}
+\w{T}_{d(a}\ep_{b)c}{}{}^d\,.
\label{s13}\end{equation}
If $M_{abc}=M_{[ab]c}$, then (\ref{s8}) -- (\ref{s9}) give
\begin{equation}
M_{abc}\equiv M_{[ab]c}=S_{[a}h_{b]c}+\w{S}_{dc}\ep_{ab}{}{}^d
+M\ep_{abc}\,,
\label{s14}\end{equation}
where $S_a=M_{ab}{}{}^b$.

Equations (\ref{s12}) -- (\ref{s14}) give the complete covariant 
decomposition of any rank--3 spatial tensor. An example of (\ref{s14})
is provided by the decomposition of the commutators $\gamma^c{}_{ab}$
of an orthonormal triad \cite{em}. 
The decomposition (\ref{s13}) applies to $\D_cE_{ab}$, $\D_cH_{ab}$
and $\D_c\sigma_{ab}$.
In the case of $\D_cE_{ab}$, 
contracting (\ref{s13}) with $h^{bc}$ gives
$\D^b E_{ab}={\ts{3\over2}}M_a$,
while contraction with $\ep^{cb}{}{}_e$ and symmetrisation on
$ea$ gives
$\c E_{ea}=-{\ts{3\over2}}\w{T}_{ea}$.
Thus (\ref{2}) is established. The corresponding relations for
$H_{ab}$ and $\sigma_{ab}$ follow immediately.

\section{Gravitational waves}

The identification of the distortions as the parts of the 
Weyl derivative that are uncoupled from the Bianchi identities,
suggests that they play an important role in the existence of
gravitational waves. 
We show here that this is indeed the case.

Gravitational waves in cosmology are gauge--invariant 
transverse tensor perturbations of the homogeneous and isotropic 
FRW spacetime. 
The covariant characterisation of gravitational waves in terms of
the Weyl tensor was introduced by Hawking \cite{h} and developed
in the
covariant and gauge--invariant perturbation theory 
of Ellis and Bruni \cite{bruell93}. In this approach,
the fundamental equations are the covariant propagation and
constraint equations (\ref{a1}) -- (\ref{a12}) given in the
Appendix, with the right hand sides set to zero (i.e. linearised).
Gravitational waves are described by $E_{ab}$ and $H_{ab}$, 
governed by the linearised form of the
Bianchi identities (\ref{a9}) --
(\ref{a12}). In the
absence of scalar and vector modes, the gradients of $\rho$, $p$ and
$\Theta$ are zero, together with $\dot{u}_a$ and $\omega_a$. It 
follows from the linearised version of
equations (\ref{a4}), (\ref{a11}) and (\ref{a12})
that $\sigma_{ab}$, $E_{ab}$ and $H_{ab}$ 
are divergence--free. These tensors satisfy wave equations 
\cite{h}, \cite{hogell96} provided
their curls are non--zero (strictly speaking, the curl of
the curl must be non--zero). For example, taking the time derivative
of the linearisation of equation (\ref{a10}), using the
linearisation of equations (\ref{a1}), (\ref{a6}), (\ref{a9}) and
the identities (\ref{a14}) and (\ref{a15}), we get
\begin{eqnarray}
\Box^2H_{ab} &\equiv& -\ddot{H}_{ab}+\D^2H_{ab} \nonumber\\
&=& {\ts{7\over3}}\Theta\dot{H}_{ab}
+\left({\ts{2\over3}}\Theta^2-2p\right)H_{ab}\,,
\label{hwe}
\end{eqnarray}
in agreement with \cite{hogell96} and \cite{dbe}.

The derivation of (\ref{hwe}) reflects the
known result \cite{hogell96} that a necessary covariant condition for
gravitational waves in cosmology is
\begin{equation}
\D^bE_{ab}=0\neq\c E_{ab}\,,\quad \D^bH_{ab}=0\neq\c H_{ab}\,.
\label{gw}\end{equation}
Now the identity (\ref{lap}) derived in Section 2 for the covariant 
Laplacian allows us to refine this result. In the linearised case, the
term ${\cal N}[H]_{ab}$ given by (\ref{lapn}) vanishes, and we have
\begin{equation}
\D^2H_{ab}=3\D^c\w{H}_{cab}-2\left(\rho-{\ts{1\over3}}\Theta^2\right)
H_{ab}\,,
\label{hwe2}\end{equation}
since the divergence is zero for gravitational waves. It follows
that if the distortion of $H_{ab}$ vanishes (strictly, if the
divergence of the distortion vanishes), then $H_{ab}$ satisfies a
covariant Helmholtz equation
\begin{equation}
\D^2H_{ab}=-\left({6{\cal K}\over a^2}\right)H_{ab}\,,
\label{hwe3}\end{equation}
where we have used the background Friedmann equation to evaluate
the coefficient in (\ref{hwe2}), with $a$ the scale factor
and ${\cal K}=0,\pm1$ the spatial curvature of the background.
For ${\cal K}=0$, the Laplacian of $H_{ab}$ vanishes and there
can be no wave propagation. For ${\cal K}\neq0$,
$H_{ab}$ is proportional to
a tensor eigenfunction
of the covariant Laplacian. These covariant
eigenfunctions satisfy \cite{h}
\[
\D^2 Q^{(k)}_{ab}=-{k^2\over a^2}Q^{(k)}_{ab}\,,\quad
Q^{(k)}_{ab}=Q^{(k)}_{\langle ab\rangle }\,,
~\D^bQ^{(k)}_{ab}=0=\dot{Q}^{(k)}_{ab}\,,
\]
where $k$ determines the comoving wave number.
For ${\cal K}=-1$, (\ref{hwe3}) shows that
there is no real value of $k$ (and thus no smooth 
eigenfunction), which forces $H_{ab}=0$, so that there
are no gravitational waves. For ${\cal K}=+1$, (\ref{hwe3}) implies
$k=\sqrt{6}$, so that there is at most one wavelength admitted. 
However, this rules out general gravitational waves.
A similar result follows if
we start with zero distortion of $E_{ab}$.
Thus we have the
extension of (\ref{gw}):\\
{\em a necessary covariant condition for the existence of
gravitational waves in cosmology is}
\begin{eqnarray}
\D^bH_{ab}=0\,,\quad \c H_{ab} \neq 0 \neq \D_{\langle c}
H_{ab\rangle }\,,
\nonumber\\
\D^bE_{ab}=0\,,\quad \c E_{ab} \neq 0 \neq \D_{\langle c}
E_{ab\rangle }\,.
\label{gw2}\end{eqnarray}

\section{Conclusion}

We have found the complete covariant decomposition of the
derivatives of the electric and magnetic parts of the
Weyl tensor. This extends the
covariant characterisation of the locally free gravitational field
to its space--time gradient.
The derivatives decompose into time derivatives and spatial
curls, divergences and distortions (totally symmetric
trace-free parts). 
The covariant Ricci and 
Bianchi equations given in the Appendix involve all these
parts of $\nabla_cE_{ab}$ and $\nabla_cH_{ab}$
except for the distortions
$\w{E}_{cab}\equiv\D_{\langle c}E_{ab\rangle }$ and 
$\w{H}_{cab}\equiv\D_{\langle c}H_{ab\rangle }$, 
which do not occur in the
equations. They may be thought of as the uncoupled parts of
the derivatives, containing at most $2\times7=14$ 
independent components out of the maximal
$40$ independent components of $\nabla_eC_{abcd}$.
Together with the curls, the distortions are crucial for
the existence of gravitational waves.

At any spacetime event ${\cal P}$, specification of $C_{abcd}$ and 
$\nabla_eC_{abcd}$ determines a first--order Taylor
approximation to $C_{abcd}$  near ${\cal P}$.
The divergences are determined algebraically 
at ${\cal P}$ via the Bianchi constraints (\ref{a11}) and 
(\ref{a12}). The curls algebraically determine the
time derivatives via the Bianchi propagation equations (\ref{a9})
and (\ref{a10}), so determining the future evolution of the system. 
Together with the distortions at ${\cal P}$, they are free data 
at that point, in that they are not determined by the matter there
(i.e. by $\rho$, $p$, $u^a$ and their first derivatives).
Rather their values there are determined non--locally, their time
evolutions along the fluid flow--line through ${\cal P}$ being
determined by the free data at earlier times (through spatial 
derivatives of the Bianchi propagation equations (\ref{a9}) and 
(\ref{a10}) for $E_{ab}$ and $H_{ab}$), and their spatial derivatives 
being restricted by spatial derivatives 
of the  Bianchi constraint equations (\ref{a11}) and (\ref{a12}). 

It is clear from this that the distortions at ${\cal P}$ are
the least coupled
parts of the
derivative of the Weyl tensor. Together with the 
curls, they are the free data that can be given at a point.
Since the curls contain at most $2\times5=10$ independent components,
there are at ${\cal P}$ in general $14+10=24$ unconstrained
components of $\nabla_eC_{abcd}$. These tensor properties at a point 
are reminiscent of Penrose's result \cite{p}
that the totally symmetrised derivatives of the Weyl spinor are
freely specifiable at a point in the vacuum case (see
\cite{ma} for the non--vacuum generalisation, and \cite{at} for
further discussion and application, of Penrose's result).

We have shown how the distortions of $H_{ab}$ and $E_{ab}$
must be non--vanishing if there
are to be gravitational waves in cosmology, thus 
refining the known covariant condition that the curls must be 
non--zero. Further investigation of the role of the distortions
is clearly warranted. In particular,
the covariant classification of cosmological
solutions via $E_{ab}$, $H_{ab}$ and their divergences and curls 
(see \cite{ell71}, \cite{matetal94}, \cite{veugleel96}, 
\cite{maart96}, \cite{mle}, \cite{ee}), could be extended by
considering also the distortions.

Finally, we note that 
the electromagnetic analogue of these results
follows from the covariant decomposition 
$$
\D_bE_a=\D_{\langle b}E_{a\rangle }+{\ts{1\over3}}\D^cE_ch_{ab}
-{\ts{1\over2}}\c E_c\ep_{ab}{}{}^c \,,
$$
and similarly for the magnetic field vector $H_a$. This shows that
the distortions of $E_a$ and $H_a$                      
are `free' in the sense that they are not directly
constrained by Maxwell's equations (see \cite{ell73}). Together
with the curls, these distortions, as in the tensor case, are
crucial for the existence of electromagnetic waves.

\newpage

\appendix
\section{Covariant Ricci and Bianchi equations}

The Ricci identity for $u^a$ and the Bianchi identities 
(incorporating the field equations via the Ricci tensor), 
are covariantly split into propagation and constraint equations 
in \cite{ell71}. For a perfect fluid source, these equations 
are in our notation the following:\\

Ricci:
\begin{eqnarray}
\dot{\Theta}+{\ts{1\over3}}\Theta^2-\D^a\dot{u}_a
+{\ts{1\over2}}(\rho+3p)
&=&-\sigma_{ab}\sigma^{ab}
+2\omega_a\omega^a +\dot{u}_a\dot{u}^a\,, \label{a1}\\
\dot{\omega}_{\langle a\rangle }+{\ts{2\over3}}\Theta\omega_a
+{\ts{1\over2}}\c \dot{u}_a&=&\sigma_{ab}\omega^b\,,\label{a2}\\
\dot{\sigma}_{\langle ab\rangle }+{\ts{2\over3}}\Theta\sigma_{ab}
-\D_{\langle a}\dot{u}_{b\rangle }+E_{ab} &=&
-\sigma_{c\langle a}\sigma_{b\rangle }{}^c \nonumber\\
&& -\omega_{\langle a}\omega_{b\rangle }
+\dot{u}_{\langle a}\dot{u}_{b\rangle }\,,\label{a3}\\
{\ts{2\over3}}\D_a\Theta-\D^b\sigma_{ab}+\c\omega_a &=&
2\ep_{abc}\omega^b\dot{u}^c\,,\label{a4}\\
\D^a\omega_a &=& \dot{u}^a\omega_a\,,\label{a5}\\
H_{ab}-\c\sigma_{ab}-\D_{\langle a}\omega_{b\rangle } &=& 
2\dot{u}_{\langle a}
\omega_{b\rangle }\,,\label{a6}
\end{eqnarray}\\

Bianchi:
\begin{eqnarray}
\dot{\rho}+(\rho+p)\Theta &=& 0\,,\label{a7}\\
(\rho+p)\dot{u}_a+\D_ap &=& 0\,,\label{a8}\\
\dot{E}_{\langle ab\rangle }+\Theta E_{ab}
+{\ts{1\over2}}(\rho+p)\sigma_{ab}-\c H_{ab} &=&
3\sigma_{c\langle a}E_{b\rangle }{}^c 
 +\omega^c \ep_{cd(a}E_{b)}{}^d\nonumber\\
&&-2\dot{u}^c\ep_{cd(a}H_{b)}{}^d\,,\label{a9}\\
\dot{H}_{\langle ab\rangle }+\Theta H_{ab}
+\c E_{ab} &=&
3\sigma_{c\langle a}H_{b\rangle }{}^c
+\omega^c \ep_{cd(a}H_{b)}{}^d \nonumber\\
&&+2\dot{u}^c\ep_{cd(a}E_{b)}{}^d\,,\label{a10}\\
\D^bE_{ab}-{\ts{1\over3}}\D_a\rho &=& \ep_{abc}\sigma^b{}_dH^{cd}
-3H_{ab}\omega^b\,,\label{a11}\\
\D^bH_{ab}-(\rho+p)\omega_a &=&-\ep_{abc}\sigma^b{}_dE^{cd}
+3E_{ab}\omega^b\,.\label{a12}
\end{eqnarray}\\

These equations generalise those given for 
the case of irrotational dust in \cite{maart96}.
In the case of small anisotropies and inhomogeneities, i.e. when
the cosmology is almost FRW, the right hand sides of the above
equations can be set to zero. The resulting 
linearised equations are the
basis for covariant and gauge--invariant perturbation theory
\cite{bruell93}.

By projecting the Ricci identity for rank--2 tensors,
we can generalise the identities in \cite{maart96} for the commutation
of covariant time and spatial derivatives, and for the curl of the
curl:\\

Commutation of spatial derivatives (exact, non--linear):
\begin{eqnarray}
\D_{[a}\D_{b]}S^{cd} &=& {\ts{2\over3}}\left({\ts{1\over3}}\Theta^2-
\rho\right)S_{[a}{}^{(c}h_{b]}{}^{d)}-2S_{[a}{}^{(c}\left\{
E_{b]}{}^{d)}-{\ts{1\over3}}\Theta\sigma_{b]}{}^{d)}\right\}
\nonumber\\
&& +2h_{[a}{}^{(c}\left\{E_{[b]e}-{\ts{1\over3}}\Theta\sigma_{b]e}
\right\}S^{d)e}-2\sigma_{[a}{}^{(c}\sigma_{b]e}S^{d)e} \nonumber\\
&& -2S_{[a}{}^{(c}\left\{\omega_{b]}\omega^{d)}-\omega_e\omega^e
h_{b]}{}^{d)}+{\ts{1\over3}}\Theta\ep_{b]}{}^{d)}{}{}_e\omega^e
\right\} \nonumber\\
&&+2\left\{\sigma_{[a}{}^{(c}\ep_{b]ef}+\sigma_{e[b}\ep_{a]}{}^
{(c}{}{}_f\right\}\omega^fS^{d)e} \nonumber\\
&& +h_{[a}{}^{(c}\left\{\omega_{b]}\omega_e+{\ts{1\over3}}
\Theta\ep_{b]ef}\omega^f\right\}S^{d)e} 
-\ep_{abe}\omega^e\dot{S}^{\langle cd\rangle }\,.
\label{a13}
\end{eqnarray}\\

Commutation of curl and time derivative (linearised):
\begin{equation}
\left(\c S_{ab}\right)^{\rd}=\c\dot{S}_{ab}-{\ts{1\over3}}\Theta
\c S_{ab}\,.
\label{a14}\end{equation}

Curl of curl (linearised):
\begin{equation}
\c\c S_{ab}=-\D^2S_{ab}+{\ts{3\over2}}\D_{\langle a}\D^cS_{b\rangle c}
+\left(\rho-{\ts{1\over3}}\Theta^2\right)S_{ab}\,,
\label{a15}\end{equation}
where $\D^2=\D^a\D_a$ is the covariant Laplacian.


\newpage

\section*{References}

\begin{thebibliography}{99}

\bibitem{ell71}
Ellis G F R  1971 Relativistic cosmology
{\em General Relativity and Cosmology} ed R K Sachs 
(New York: Academic Press) pp 104--179
	
\bibitem{ell73}
Ellis G F R 1973 Relativistic cosmology
{\em Cargese Lectures in Physics, Vol. VI} ed E Schatzmann 
(New York: Gordon and Breach) pp 1--60

\bibitem{bruell93}
Ellis G F R and Bruni M 1989 A covariant and gauge--free approach to 
 density fluctuations in cosmology {\em Phys. Rev.} D {\bf 40} 
 1804 

\bibitem{matetal94}
Matarrese S, Pantano O and Saez D 1994 General relativistic
dynamics of irrotational dust: Cosmological implications {\em
Phys. Rev. Lett.} {\bf 72} 320

\bibitem{veugleel96}
van Elst H, Uggla C, Lesame W M, Ellis G F R and Maartens R 1997
Integrability of 
irrotational silent cosmological models {\em Class. Quantum Grav.} 
to appear

\bibitem{h}
Hawking S W 1966 Perturbations of an expanding universe 
{\em Astrophys. J.} {\bf 145} 544

\bibitem{hogell96}
Hogan P and Ellis G F R 1997 Propagation of information by 
electromagnetic and gravitational waves in cosmology
{\em Class. Quantum Grav.} {\bf 14} A171

\bibitem{ellhog96}
Ellis G F R and Hogan P 1997 The electromagnetic analogue of
some gravitational perturbations in cosmology {\it Gen. Rel. Grav.}
{\bf 29} 235

\bibitem{maart96}
Maartens R 1997 Linearisation instability of gravity waves?
{\it Phys. Rev.} D {\bf 55} 463

\bibitem{mle}
Maartens R, Lesame W M and Ellis G F R 1997 Consistency of dust
solutions with div $H=0$ {\em Phys. Rev.} D {\bf 55} in press

\bibitem{dbe}
Dunsby P K S, Bassett B A B and Ellis G F R 1997 Covariant analysis
of gravitational waves in a cosmological context {\em Class. Quantum
Grav.} to appear

\bibitem{sik96}
Siklos S T C 1996 Counting solutions of Einstein's equations
{\it Class. Quantum Grav.} {\bf 13} 1931

\bibitem{em} Ellis G F R and MacCallum M A H 1969 A class of 
homogeneous 
cosmological models {\em Commun. Math. Phys.} {\bf 12} 108

\bibitem{p}
Penrose R 1960 A spinor approach to general relativity
{\em Ann. Phys.} {\bf 10} 171

\bibitem{ma}
MacCallum M A H and Aman J E 1986 Algebraically independent $n$--th
derivatives of the Riemannian curvature spinor in a general spacetime
{\em Class. Quantum Grav.} {\bf 3} 1133

\bibitem{at}
Anderson I M and Torre C G 1996 Classification of local generalized
symmetries for the vacuum Einstein equations
{\em Commun. Math. Phys.} {\bf 176} 479

\bibitem{ee}
van Elst H and Ellis G F R 1996 The covariant approach to LRS perfect
fluid spacetime geometries {\em Class. Quantum Grav.} {\bf 13} 1099

\end{thebibliography}
\end{document}